\documentclass[conference]{IEEEtran}
\usepackage{cite}
\usepackage{graphicx}
\usepackage{amsmath,amssymb,amsfonts}
\usepackage{algorithmic}
\usepackage{textcomp}
\usepackage{xcolor}
\usepackage{booktabs}
\usepackage{stfloats}  
\usepackage{hyperref}
\usepackage{tabularx}
\usepackage{array}
\usepackage{placeins}
\usepackage[normalem]{ulem}
\usepackage[table]{xcolor}
\usepackage{array}
\usepackage{tabularx}
\usepackage{makecell}
\usepackage{enumitem}
\usepackage{booktabs}

\begin{document}

\newcommand{\vivian}[1]{{\color{purple} Vivian: #1}}
\newcommand{\billy}[1]{{\color{green} Billy: #1}} 
\definecolor{aryanbrown}{HTML}{8B5A2B}
\DeclareRobustCommand{\aryanedit}[1]{{\color{aryanbrown}\textit{#1}}}
\DeclareRobustCommand{\aryanremove}[1]{{\color{aryanbrown}\sout{#1}}}

\definecolor{CinderellaBlue}{HTML}{DCE8F6}
\definecolor{CinderellaBlueDark}{HTML}{C7D9EE}
\definecolor{CinderellaBlueText}{HTML}{2F4A6D}
\definecolor{CinderellaBlue}{HTML}{DCE8F6}
\definecolor{CinderellaBlueText}{HTML}{2F4A6D}
\definecolor{CinderellaBluePale}{HTML}{EEF4FB}
\definecolor{CinderellaBluePale}{HTML}{F4F8FD}
\definecolor{CinderellaBlueVeryPale}{HTML}{FAFCFF}

\title{\huge Rasoi: Visual Diffs for Evaluating Structural Changes Made by Coding Agents}

\author{
\begingroup
\fontsize{10.5pt}{12pt}\selectfont
\setlength{\tabcolsep}{18pt}
\renewcommand{\arraystretch}{1.08}
\begin{tabular}{ccc}
\textbf{Riya Sahni} &
\textbf{Aryan Kaul} &
\textbf{Anushka Bhatnagar} \\
Columbia University &
Columbia University &
Columbia University \\
New York City, USA &
New York City, USA &
New York City, USA \\
riya.sahni@cs.columbia.edu &
aryan.kaul@columbia.edu &
anushka.bhatnagar@columbia.edu \\[0.9em]

\textbf{Xuanming Zhang} &
\textbf{Vivian Liu} &
\textbf{Reya Vir} \\
Columbia University &
Columbia University &
Columbia University \\
New York City, USA &
New York City, USA &
New York City, USA \\
billyzhang@cs.columbia.edu &
vivian@cs.columbia.edu &
reyavir@cs.columbia.edu \\[0.9em]

& \textbf{Lydia B. Chilton} & \\
& Columbia University & \\
& New York City, USA & \\
& chilton@cs.columbia.edu &
\end{tabular}
\endgroup
}


\maketitle

\begin{abstract}
Coding agents let developers modify software rapidly through natural language. However, as agents take on larger changes, developers can lose track of how the codebase is evolving. These edits may look correct on the surface while restructuring the system in unexpected ways: duplicating logic, changing modules outside the requested scope, removing functionality developers meant to keep, or introducing architectural drift. We investigate how visual diffs can help developers understand and evaluate the structural changes that coding agents make. We present Rasoi, a design probe that maps how a codebase's architecture evolves after each agent-made edit, letting developers inspect which architectural elements and relationships were added, modified, or removed across multiple levels of abstraction. We deployed Rasoi in a week-long study with 12 developers working on their own codebases across varied technical domains. We found that visual diffs helped developers see whether changes stayed within scope, notice unexpected restructuring, and understand how modified parts related to the system as a whole. Our findings suggest that the value of oversight tools for coding agents lies less in making developers trust the agent and more in helping developers feel confident in their own evaluation of agent-made changes. We discuss implications for visual tools that support code comprehension and human oversight in AI-assisted software development.
\end{abstract}


\section{Introduction}

\begin{figure*}[t]
    \centering
    \includegraphics[width=\textwidth]{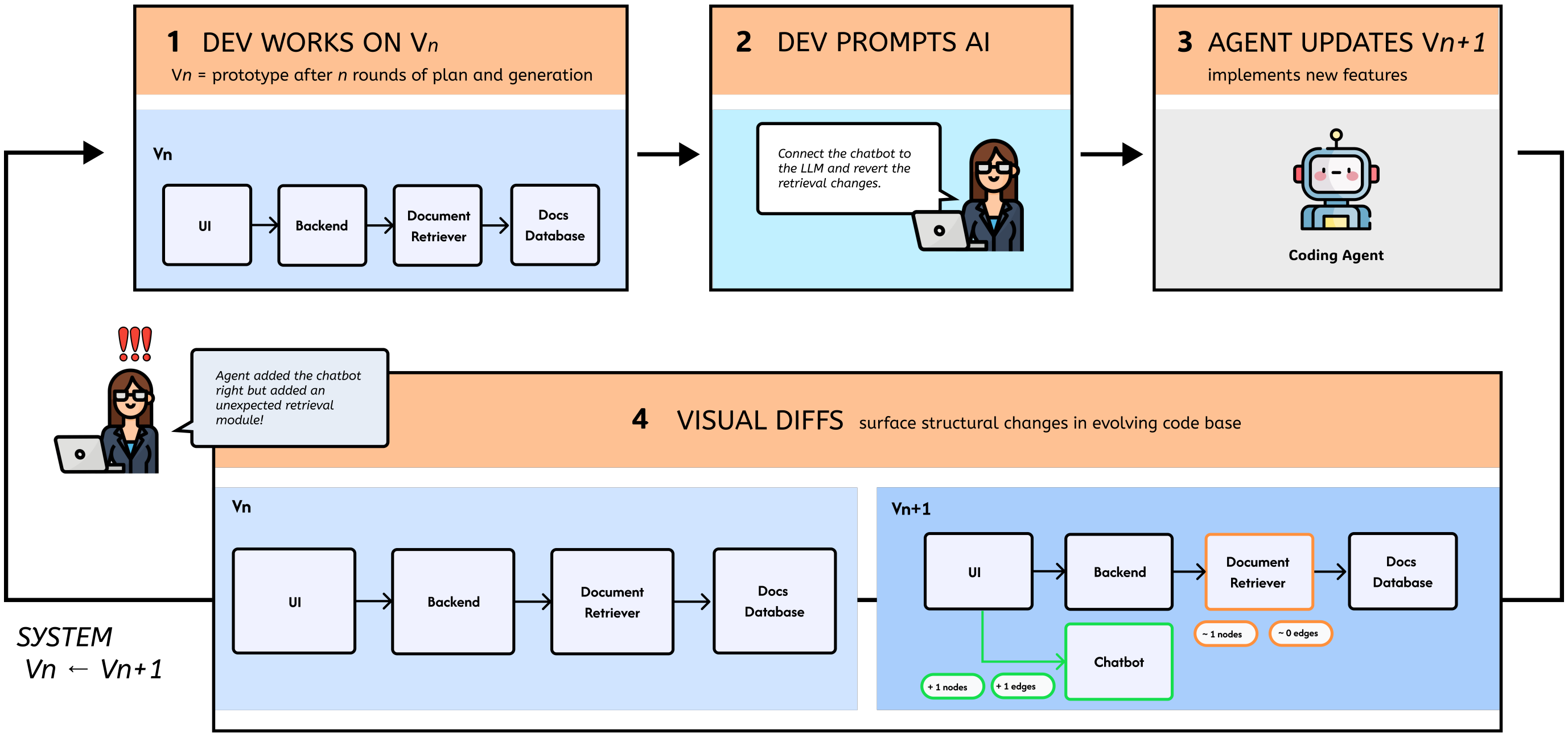}
    \caption{Rasoi helps developers inspect and evaluate structural changes made by coding agents as a codebase evolves from one version ($V_n$) to the next ($V_{n+1}$). After an agent updates the codebase, Rasoi surfaces a visual diff between $V_n$ and $V_{n+1}$, showing which architectural elements and relationships were added, modified, removed, or unexpectedly affected. After this comparison, $V_{n+1}$ becomes the current version for the next agent-made change.}
    \label{fig:teaser}
\end{figure*}

Software development is entering a new era of human-AI interaction. With coding agents like Claude Code, Cursor, and Codex~\cite{claude-code,cursor,codex}, developers can rapidly build, modify, and debug software through natural language. Software tools have long let developers work at higher levels of abstraction, from compilers and high-level languages to modern programming environments~\cite{brooks1987nosilverbullet}. Coding agents extend this shift, moving developers' attention from individual lines of code toward how codebases fit together, including how modules connect, how data flows, and how parts of a system depend on one another. As agents take on larger changes that reorganize modules, revise workflows, or rewire dependencies, developers must reason about more than whether a feature works.

These larger changes make it harder for developers to tell whether a working feature also reshaped the codebase in ways they did not intend. A coding agent may complete a requested task while also duplicating logic, introducing unnecessary dependencies, or changing connections elsewhere in the codebase. These changes might not necessarily break the feature, but they can quietly reshape the surrounding system and make future agent edits more confusing or costly. The burden falls on developers to review not just whether the feature works, but whether the codebase-level changes around the feature are ones they want to accept.

Historically, developers have used software visualizations to understand the structure of a codebase.
These representations abstract away lines of code to show how components, modules, and services relate to one another~\cite{uml,kruchten1995fourplusone,clements2002views}.
Many UML-style diagrams capture this structure at a single level, while other models, like Brown's C4, organize an evolving codebase across multiple levels of abstraction, from broad system context down to containers, components, and code-level detail~\cite{brownC4}.
C4 is especially useful because change can happen at any of these levels: a single edit can alter the overall workflow of a system or change which parts of a codebase connect to one another.

However, coding agents introduce a new challenge.
Developers now have to evaluate structural changes made by coding agents.
We use \textit{structural changes} to refer to modifications that alter the organization of a system’s architectural elements or the relationships among them, including components, modules, services, dependencies, and data- or control-flow connections. This definition is informed by prior characterizations of architecture as comprising system elements, their relationships, and the design decisions that organize them~\cite{shahbazian2017uncovering,iso42010}.
This definition raises a different question about software visualization: how can we help developers see what a coding agent changed, and whether the system is structured as intended?



To explore how visualizations might help developers understand and evaluate agent-made structural changes, we developed Rasoi, a design probe for reviewing how a codebase evolves as a coding agent changes it. When a coding agent changes the codebase, Rasoi generates a \textit{visual diff}, a map of the codebase architecture that highlights what was added, modified, or removed between versions. We build on C4 because its abstraction levels give Rasoi a repeatable set of views that can be regenerated after each agent edit and compared across versions.
We investigate two research questions. \textbf{RQ1:} How do developers use \textit{visual diffs} to understand how a codebase has changed after an agent edit? \textbf{RQ2:} When do developers need high-level architectural views versus more detailed code-level views to evaluate agent-made changes?

Our approach consisted of three parts. We first conducted formative interviews with developers who regularly use coding agents to understand what makes feature addition and bug tracing difficult in agent-assisted coding. Next, guided by these findings, we designed Rasoi around design goals for making agent-made structural change visible and inspectable.
Finally, we deployed Rasoi in a 1-week longitudinal probe study with 12 developers using it on their own codebases while working with coding agents.



This paper makes three contributions:

\begin{enumerate}
    \item \textbf{Rasoi, a design probe}  that uses visual diffs and drill-down across abstraction levels to support inspection of structural changes made by coding agents to a codebase.
    \item \textbf{Empirical findings from a 1-week longitudinal deployment with 12 developers} showing that visual diffs helped developers verify whether changes stayed within their intended scope, surface unexpected side effects and regressions, and understand how modified parts of a codebase related to the rest of the system.

    \item \textbf{Design implications for visualizations in agent-assisted coding:} visual support should help developers review agent-made changes for themselves, provide the right level of detail for the developer’s current question, and appear at moments of evaluation rather than continuously.
\end{enumerate}


\section{Related Work}

Rasoi builds on three lines of work: foundational research on how developers understand and modify evolving software systems, studies of oversight and validation in AI-assisted programming, and recent systems that make generated code, traces, and design decisions more inspectable.

\subsection{Foundations in Program Comprehension, Mental Models, and Software Structure}
Long before coding agents, researchers studied how developers understand, modify, and recover structure from software systems. Prior work describes programmers' mental models as organized representations of what they know about programs, tasks, and concepts \cite{heinonen2023mentalmodels}. These mental models span both implementation-level details and architecture-level design decisions. Developers need implementation-level knowledge about control flow, data flow, and how features are encoded, but they also need architecture-level knowledge about the major parts of a system \cite{heinonen2023mentalmodels,sillito2008askinganswering,sillito2008asking}. During software evolution tasks, programmers ask many kinds of questions to recover this understanding, including where behavior lives, how components relate, and what other parts of the code might be affected by a change \cite{sillito2008askinganswering,sillito2008asking}.

One response to this problem has been to give developers external representations of program structure. Software reflexion models compare a developer's high-level model of a system against the structure recovered from source code, making visible where that model aligns with or diverges from the implementation \cite{murphy1995reflexion,murphy2001reflexion}. Other systems support comprehension by preserving spatial context, arranging related code fragments, or surfacing contextual annotations near the code itself \cite{dealwis2006visualmomentum,bragdon2010codebubbles,adeli2020annotations}. Software architecture frameworks offer another way to move between overview and detail. The C4 model, for example, organizes a system across context, containers, components, and code \cite{brownC4}. Together, this work gives developers ways to recover structure that is otherwise scattered across source code, documentation, and navigation.

Agent-assisted development makes this long-standing comprehension problem more urgent. Coding agents can generate and revise code quickly, often across multiple files and components, which shifts developers' work toward reviewing, validating, and repairing changes they did not author line by line \cite{mozannar2024reading,liang2024usability,tang2024validating}. These edits may also come with less of the context that helps developers make sense of human-authored changes. A developer may see what changed, but not why a dependency moved, why a previously working behavior disappeared, or what design assumption led the agent to make the change. Rasoi builds on foundational work in mental models, architecture views, and change comparison, but applies it to a newer problem: maintaining structural understanding while a coding agent repeatedly changes the codebase.

\subsection{Oversight and Validation in Agent-Assisted Programming}

Recent work on AI-assisted programming shows how program comprehension changes when developers review code generated by an assistant rather than code they wrote themselves. AI coding assistants are increasingly integrated into programming workflows, where they generate, complete, explain, and repair code through natural-language interaction~\cite{ross2023programmersassistant,liang2024usability,sergeyuk2025codingassistants}. These systems can speed up development, but they also shift developers' work toward supervising, validating, steering, and repairing code they did not author line by line~\cite{ross2023programmersassistant,mozannar2024reading,liang2024usability,tang2024validating}.

This validation work is not only about checking whether code compiles. An assistant may implement the requested feature while still making changes that do not fit the surrounding system, such as routing data through the wrong path, leaving stale dependencies, removing prior behavior, or introducing technical debt \cite{mozannar2024reading,liang2024usability,tang2024validating}. Recent studies show that developers' mental models of AI-driven code tools can diverge from how those tools behave, and that developers engage in substantial validation and repair work around LLM-generated code \cite{desolda2025mentalmodels,tang2024validating}. Together, this work shows that the central challenge is not simply generating more code. Developers need ways to understand what the assistant changed, how much of the system it touched, and whether that change should be accepted into the evolving codebase \cite{mozannar2024reading,desolda2025mentalmodels,tang2024validating}.

These studies make clear that oversight is now part of the programming workflow. But they mostly characterize the human effort of inspecting, validating, and repairing AI-generated code. They do not give developers a way to see the agent's changes as a structural object in the codebase. Rasoi fills this gap by making the footprint of an agent-made edit visible across versions. Rather than asking developers to infer scope from generated text, tests, or file-level diffs alone, Rasoi lets them inspect whether the change stayed within the scope they expected and how it reshaped the surrounding system.

\subsection{Inspectable Representations for AI-Assisted Programming}

Recent agent-era systems increasingly provide developers with inspectable representations around AI-assisted programming. Some help users shape generated artifacts before they are committed, some explain generated code after it exists, and others support debugging through program graphs, IDE context, or execution traces. Together, this work shows that developers need more than generated code or chat explanations when working with AI coding assistants.

Pail supports exploration of code generation by helping users explore higher-level formalizations of problems and solutions \cite{zamfirescu2025pail}. BISCUIT supports computational notebooks by inserting ephemeral UIs between prompts and generated code, allowing users to adjust variables and steer generation \cite{cheng2024biscuit}. These systems make the design space around generation more inspectable while the artifact is still being shaped. Ivie anchors explanations to generated code at the block and expression level \cite{yan2024ivie}, while CodeVoyager lets natural-language queries highlight regions of a call graph and generates control-flow graphs for explanation \cite{kim2026codevoyager}. These systems help developers inspect generated code or a current program representation. Other systems debug and do behavioral inspection. ROBIN gathers IDE and developer context before proposing repairs through a multi-agent extension to GitHub Copilot Chat \cite{bajpai2024fix}, while TraceDiff highlights where an incorrect execution trace begins to diverge from an expected one \cite{suzuki2017tracediff}.

Across these systems, the object of inspection is usually the generated artifact, the current code structure, or runtime behavior. Trailblazer comes closest to Rasoi because it uses an agent to inspect codebase-spanning behavior and packages discovered traces as summaries, tours, and replayable walkthroughs for developer questions \cite{yan2025trailblazer}. However, Trailblazer makes execution-time behavior navigable within a fixed version of a codebase. Rasoi focuses on a different object of inspection: how the structure of a codebase changes across agent edits.

This distinction matters because the risks developers described were often not isolated to a generated snippet or a single runtime failure. They needed to know whether an agent-made change stayed within scope, whether it touched unexpected modules, and how the surrounding architecture shifted across versions. Rasoi therefore makes \textit{structural change} across agent edits inspectable, so developers can see how those changes fit into the larger structure of the system and whether that structure stays within scope.
Rasoi brings architecture views, C4-style abstraction levels, and diff-style comparison together into an \textit{oversight surface} for agent-generated codebase evolution \cite{murphy1995reflexion,murphy2001reflexion,sillito2008asking,brownC4,suzuki2017tracediff}. We turn next to how Rasoi instantiates this oversight surface, and to findings from a week-long deployment study.
\section{Rasoi: Design Probe}

\subsection{Design Considerations}

Rasoi is a design probe for studying how developers inspect and evaluate agent-made changes in evolving codebases. \footnote{We plan to open-source Rasoi upon acceptance.} Implementation and deployment details for Rasoi are provided in Appendix~A. 
To ground Rasoi's design, we held informal formative conversations with 8 developers who regularly use coding agents in their work. Across these conversations, participants described difficulty noticing unexpected changes made elsewhere in the system, tracing bugs introduced through agent edits, and telling whether a repeatedly edited codebase still matched the workflow they had in mind. They expressed interest in diff-style visualizations that could show how a system changed over time, support movement from high-level architecture to more specific implementation details, and fit naturally into practices like code review or PR review. Guided by these recurring concerns, we designed Rasoi around three goals: (DG1) making workflow changes visible over time, (DG2) supporting movement between architectural overviews and code-level detail, and (DG3) preserving important details about the codebase without cluttering the visualization.


\textbf{DG1: Showing how architecture changes over time.}
One early question was whether Rasoi should only show what the codebase looks like now, or also show how it got there through agent-made changes.
A static view of the codebase can help developers orient themselves to the current structure of a system, but it does not reveal how that structure has evolved or whether an agent changed unexpected parts of the codebase along the way. Our formative interviews suggested that this delta was important. Developers often struggled not only with understanding the current state of a system, but also with understanding what had changed and whether the latest version still matched the workflow they thought existed. We therefore designed Rasoi around versioned visual maps that foreground how a codebase changes across agent-driven iterations, not just what the current system looks like. Each session consists of multiple snapshots that can be compared over time, allowing us to observe how developers use visual diffs to localize edits, inspect architectural drift, and reason about newly introduced bugs.

\textbf{DG2: Supporting abstracted views of a codebase.}
High-level views can help developers quickly re-establish what a system is and how its parts relate, but they often hide the details needed to trace a bug, inspect a backend step, or verify whether a specific component is still active. Detailed views provide this evidence, but they can become overwhelming if too much is shown at once. We therefore designed Rasoi to support movement across abstraction levels.

Prior software architecture frameworks offer useful ways to represent complex systems from different perspectives~\cite{kruchten1995fourplusone,clements2002views}. These approaches are well-suited for documenting different aspects of a system, such as its structure, deployment, or runtime behavior. Rasoi required a different property---a repeatable set of views that could be regenerated after each agent edit and compared across versions of the same codebase.

We adopted Brown’s C4 model because its context, container, component, and code-level views provided this repeatable structure~\cite{brownC4}. These levels allowed Rasoi to show comparable architectural maps across iterations while still supporting movement from broad system context to implementation-level evidence. They also reflected how developers in our formative conversations described navigating their codebases. Participants often began with a broad system view, then narrowed into services or modules, and finally inspected the functions or files where an agent-made change had landed. This made C4 a practical fit for Rasoi: stable enough to support version-to-version comparisons, yet granular enough to help developers move from an architectural overview to code-level inspection.


\textbf{DG3: Preserving a legible story of the system instead of rendering everything that the codebase contains.}

We had to consider how literal the visualizations should be. AI-generated architecture diagrams can easily become cluttered with labels and relationships, especially for large codebases. A more detailed visualization might capture more information, but it would also be harder for users to make sense of what the visual is showing. We therefore designed Rasoi to preserve a legible story of the system by showing only primary modules and explicit relationships across versions. This let us focus on how participants verify and make sense of an evolving codebase, rather than on whether they could decode a noisy diagram.

\textbf{Deploying the probe in participants’ real work environments.}
Finally, we had to decide whether to study Rasoi through standardized tasks or through participants’ own ongoing software work.
Standardized tasks would have made sessions between users easier to compare, but they would have also “artificialized” the thing that we were trying to study - how are developers using visualizations of their codebases while adding new features, tracing bugs, and collaborating with others in real codebases.
Because our research questions concern how mental models are maintained during agent-assisted code generation, we chose to deploy Rasoi directly on participants’ machines and ask them to use it while working on their own systems over 1 week. 
This allowed the probe to surface naturally occurring moments in which participants had to decide whether to accept an agent-made change, inspect its structural side effects, or intervene before a questionable implementation introduced bugs or technical debt.


\subsection{Rasoi Walkthrough}
\begin{figure*}[t]
    \centering
    \includegraphics[width=\textwidth]{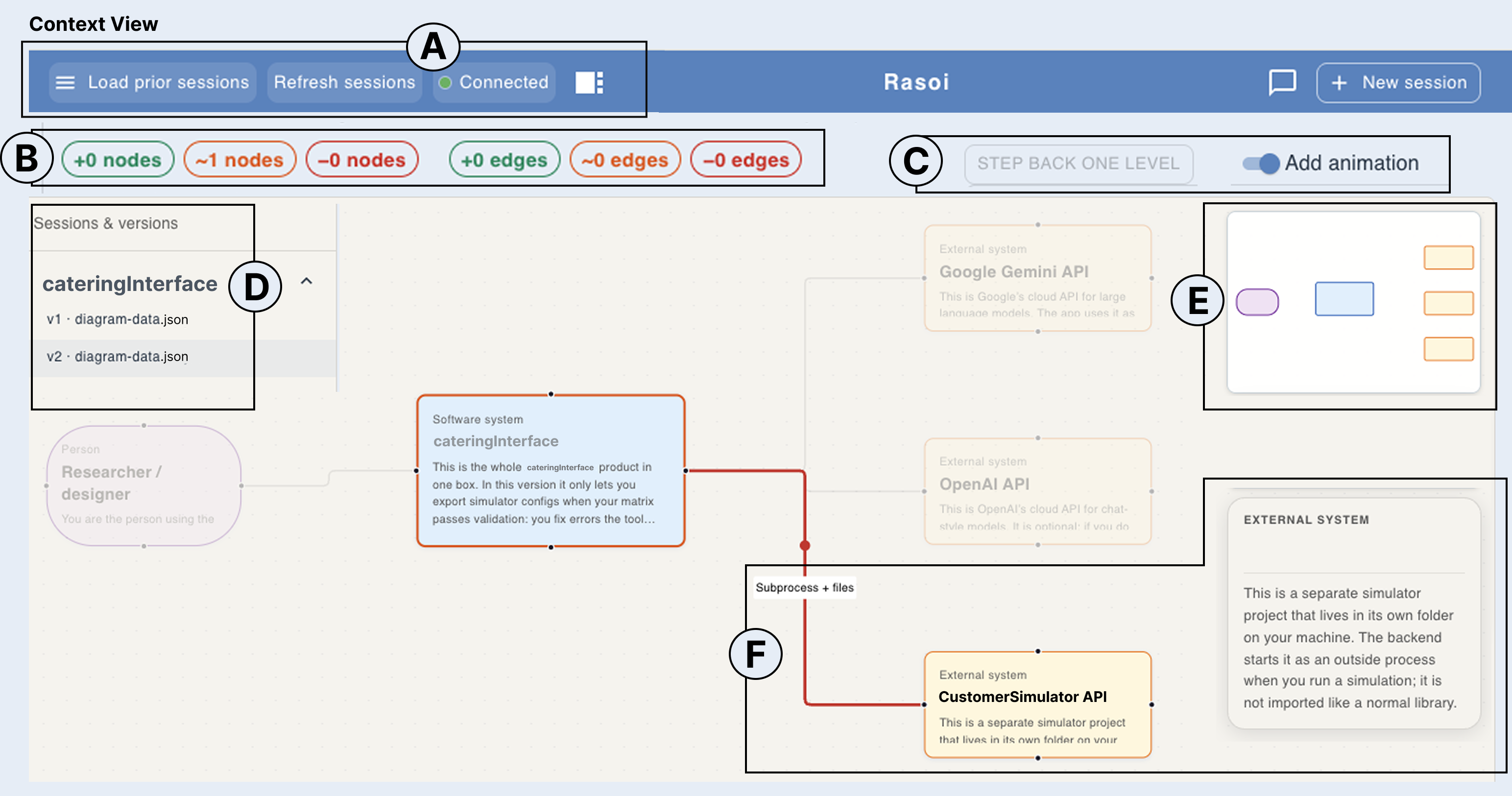}
    \caption{\textbf{Rasoi’s main interface at the context level. The interface includes session controls for loading and refreshing prior visualization sessions (A), a diff summary showing added, modified, and removed nodes and edges between snapshots (B), navigation controls for moving across abstraction levels and animating data flow (C), a session/version sidebar (D), a minimap that situates the current view within the larger system (E), and the main architecture canvas (F).}}
    \label{fig:rasoi_main}
\end{figure*}
Rasoi represents a codebase as a version-controlled architecture map. For each codebase, it stores multiple architecture snapshots within a session, allowing users to view and manage how the system changes over time (Fig.~\ref{fig:rasoi_main}A,D). To illustrate how Rasoi is used, consider Sookie, a developer working on a web application that lets users submit catering requests through a form. Sookie prompts her coding agent: \textit{“Add an email confirmation step so that when a user submits the catering request form, they get a confirmation email.”} After the agent finishes implementing the feature, Sookie connects Rasoi to her repository and generates a new snapshot of the system. Rasoi compares this snapshot against the previous one and surfaces a color-coded diff summary showing how many nodes and edges were added, modified, or removed (Fig.~\ref{fig:rasoi_main}B). This allows Sookie to quickly check whether the new feature appears to have changed only the request submission and notification parts of the system, or whether the agent also modified unrelated components.

If the diff suggests that additional modules were modified, she can inspect those changes more closely before deciding whether to keep or revise the implementation. The main canvas shows a map of the system (Fig.~\ref{fig:rasoi_main}F), beginning with a context view that shows the software system and external services it interacts with. When Sookie hovers over a node, Rasoi shows a brief description of that component. Its immediate edges are also highlighted, along with a short description of the relationship between that node and its one-step neighbors. This allows her to first inspect the change at a high level before deciding whether she needs more detail.

\begin{figure*}[tp]
    \centering
    \includegraphics[width=\textwidth]{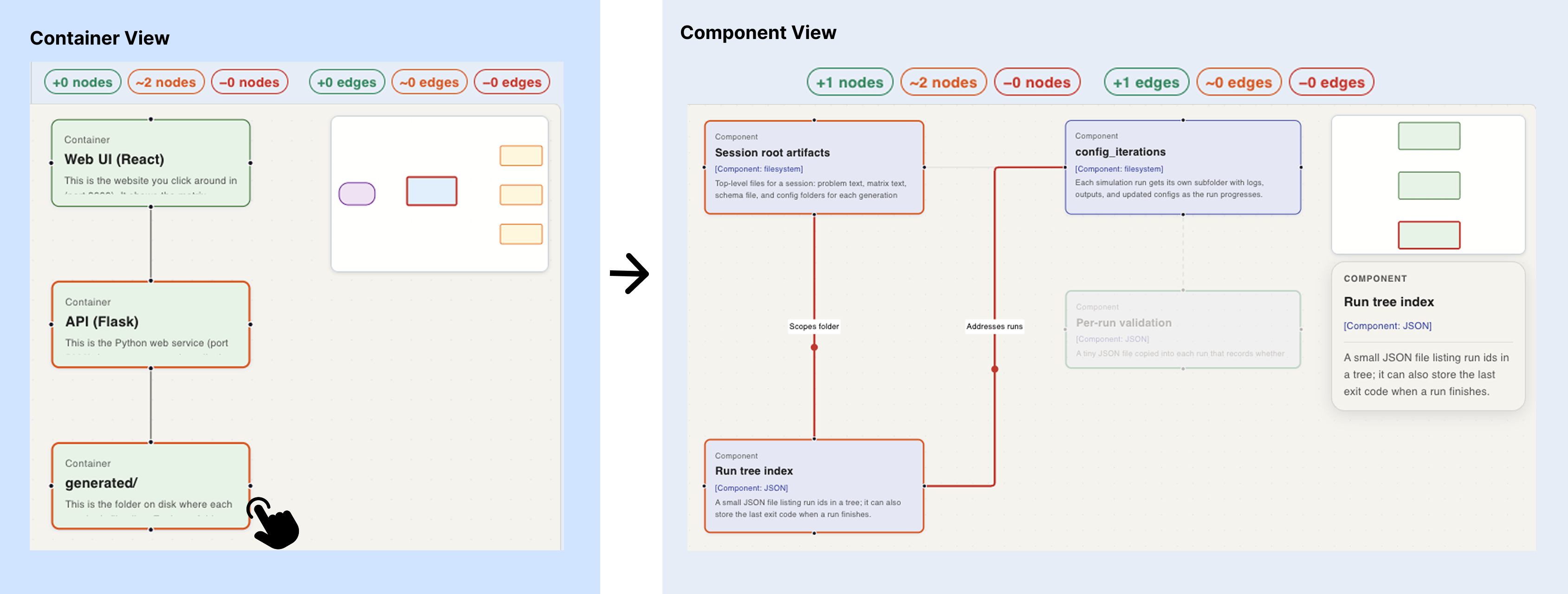}
    \caption{\textbf{Rasoi supports movement across abstraction levels.} From the context view, users can drill into container-level views (left) and then into component-level views (right), allowing them to move from broad architectural structure to more localized detail when evaluating agent-made changes.}
    \label{fig:rasoi_drilldown}
\end{figure*}

From there, Sookie can drill down into container and component views, as illustrated in Fig.~\ref{fig:rasoi_drilldown}. This allows her to move from a broad architectural overview to more granular details when she needs to inspect a specific module, backend step, or changed component. For example, if the diff suggests that the agent also modified authentication or payment-related modules, Sookie can drill into those components to check whether they were actually relevant to the feature request. The minimap situates her current view within the larger structure of the codebase (Fig.~\ref{fig:rasoi_main}E), so that even when she is examining a more granular view, she can still see how that component fits into the wider system. Finally, Sookie can move back up through abstraction levels to regain a broader view of the codebase, or animate edges to follow the direction of data flow between modules (Fig.~\ref{fig:rasoi_main}C). Prior work has shown that animation can help viewers follow movement through a visual representation~\cite{lawrence1994animations}; Rasoi uses edge animation in this spirit, helping developers trace directional relationships between changed components and the rest of the system. In this way, Rasoi supports quick visual checks after agent-made changes as well as more detailed inspection when a diff suggests that something unexpected may have happened.
\section{Longitudinal Probe Deployment}
\textbf{Participants.}
We deployed Rasoi to 12 participants who regularly used coding agents to build, modify and debug software systems. 
Participants came from a range of technical backgrounds and used coding agents in ongoing software work, including research prototypes and developer tools.
Table~\ref{tab:participants} summarizes participant backgrounds, their typical use of coding agents, and the domain/project in which they used Rasoi.
Each participant used Rasoi with a real codebase that they were actively working on. Appendix~B includes additional example diff visualizations generated by a few participants during the deployment.

\begin{table*}[bp]
\centering
\small
\setlength{\tabcolsep}{5pt}
\renewcommand{\arraystretch}{1.24}
\setlength{\abovecaptionskip}{3pt}
\setlength{\belowcaptionskip}{5pt}
\caption{Participants}
\label{tab:participants}
\begin{tabular}{
  >{\raggedright\arraybackslash}p{0.035\linewidth}
  >{\raggedright\arraybackslash}p{0.17\linewidth}
  >{\raggedright\arraybackslash}p{0.15\linewidth}
  >{\raggedright\arraybackslash}p{0.10\linewidth}
  >{\raggedright\arraybackslash}p{0.45\linewidth}
}
\toprule
\rowcolor{CinderellaBlue}
\textbf{\textcolor{CinderellaBlueText}{ID}} &
\textbf{\textcolor{CinderellaBlueText}{Role/Background}} &
\textbf{\textcolor{CinderellaBlueText}{Coding Agent}} &
\textbf{\textcolor{CinderellaBlueText}{Agent Use}} &
\textbf{\textcolor{CinderellaBlueText}{Project/Domain}} \\
\midrule
\cellcolor{CinderellaBlueVeryPale}P1 & CS PhD Student (ML) & Cursor & 3--5 days/wk & Developer documentation tool \\
\cellcolor{CinderellaBlueVeryPale}P2 & Software Engineer & Claude Code & Daily & Firebase-to-Supabase backend migration \\
\cellcolor{CinderellaBlueVeryPale}P3 & Physics PhD Student & Claude Code & 3--5 days/wk & Plasma physics simulation / data flow across modules \\
\cellcolor{CinderellaBlueVeryPale}P4 & Software Engineer & Gemini & Daily & Android accessibility service \\
\cellcolor{CinderellaBlueVeryPale}P5 & Mechanical Engineer & Cursor & 3--5 days/wk & Robotics 3D simulation software \\
\cellcolor{CinderellaBlueVeryPale}P6 & CS PhD Student (NLP) & Claude Code & 3--5 days/wk & Multi-agent simulations of human behavior \\
\cellcolor{CinderellaBlueVeryPale}P7 & CS PhD Student (ML) & Claude Code & 1--2 days/wk & ML evaluation pipeline / fine-tuning for abstention \\
\cellcolor{CinderellaBlueVeryPale}P8 & CS PhD Student (CV) & Claude Code & 1--2 days/wk & Educational web app / set theory visualizer \\
\cellcolor{CinderellaBlueVeryPale}P9 & CS MS Student & Antigravity & Daily & Portfolio web app / rendering and modularization \\
\cellcolor{CinderellaBlueVeryPale}P10 & CS PhD Student (HCI) & Cursor & 3--5 days/wk & Annotation interface web application \\
\cellcolor{CinderellaBlueVeryPale}P11 & CS PhD Student (NLP) & Claude Code & 3--5 days/wk & Agentic AI mortgage platform \\
\cellcolor{CinderellaBlueVeryPale}P12 & CS PhD Student (NLP) & Claude Code & Daily & Model training pipeline / reward design \\
\bottomrule
\end{tabular}
\end{table*}

\textbf{Procedure.}
We conducted a 7-day longitudinal probe deployment in which participants installed Rasoi locally and used it while continuing their normal agent-assisted coding work. On day 1, we onboarded participants by connecting Rasoi to their codebases and showing them how to use the interface, including how to load prior visualizations, navigate across abstraction levels, and inspect architectural diffs. Participants then used Rasoi during their own software work.

\textbf{Data collection.}
We collected three forms of data during the deployment.
First, we checked in with participants each day and asked a lightweight reflection question: ``What was your experience using Rasoi today?'' We used follow-up prompts when needed to understand what task participants were working on, what part of the codebase they were trying to reason about, which Rasoi views they used, and whether the visualization helped them understand or evaluate changes made by the coding agent. Second, we conducted a post-study interview with each participant about their overall use of Rasoi, including which views they found most helpful, when they preferred higher-level overviews to lower-level views, and how Rasoi fit into their existing workflows. Third, we collected probe artifacts, including the versioned visualizations that participants created or loaded during the study.

\textbf{Analysis.} Two researchers led the qualitative analysis of participants' daily reflections, post-study interviews, survey responses, and generated visual artifacts, working through the data iteratively to identify recurring patterns. They focused on moments when participants used Rasoi to localize change, compare versions, navigate abstraction levels, trace bugs, or explain a codebase to others. Five researchers then reviewed the emerging themes across paper meetings, refining interpretations and selecting representative quotes and artifacts. We organized the final themes around the research questions and design goals. Survey responses were used to contextualize and triangulate the qualitative findings.




\section{Findings}

\subsection{How do developers use visual diffs to understand how a codebase has changed after an agent edit?}

\subsubsection{Visual diffs helped developers verify that agent-made changes stayed within intended scope} Several participants used Rasoi to check whether an agent had changed only the parts of the codebase they expected. P1 used Rasoi to confirm that “there was nothing new that was added, which is good because that’s kind of what I expected, and nothing was ruined,” then drilled down in the visual to confirm that “the command handler, like the CLI stuff that was changed, [was] expected,” so the visualization “helped verify that the agent was only touching the parts of the code that should have been changing.” 
P1 described this as part of their normal workflow: after a feature seemed correct, they would inspect the visual that was generated, revise the code if needed, and then commit. The same kind of checking appeared in larger structural changes that participants made to their codebases. During a refactor, P4 said the visualization showed that “the flow still held together after the refactor,” and that seeing the modified files in the diagram “gained my trust.” 

P8 described a similar use after asking an agent to add a specific visualization to an existing educational web application. Looking at the diff, they said it ``exactly points out that [the agent] has added a visualizer for the exact same thing,” and that while a new module had been added, “only 3 out of the 4 were modified” to make that new component work. The diff helped verify that the agent had changed only the expected parts of the system.

Rasoi was also useful when participants suspected that an expected structural change had not actually happened. After asking an agent to modularize a frontend-heavy codebase into cleaner parts, P9 opened Rasoi before even running the generated code and noticed that “no new nodes [had] been added or deleted, even at the lowest level of abstraction,” which felt wrong because they had “just asked [Cursor] to create a bunch of new modules.”
That lack of visible change made them skeptical, and when they ran the code, the modularization had in fact broken it (Appendix, Figure ~\ref{fig:appendix_visualizations}: A). Across these cases, Rasoi helped participants inspect whether the evolving codebase actually matched what they thought the agent had done.

\newcommand{\cellbullets}[1]{%
\vspace{-1.0\baselineskip}%
  \begin{list}{\textbullet}{%
    \setlength{\topsep}{1pt}%
    \setlength{\partopsep}{0pt}%
    \setlength{\parsep}{0pt}%
    \setlength{\itemsep}{3.5pt}%
    \setlength{\leftmargin}{1.25em}%
    \setlength{\labelwidth}{0.8em}%
    \setlength{\labelsep}{0.45em}%
    \setlength{\itemindent}{0pt}%
    \setlength{\listparindent}{0pt}%
  }%
  #1
  \end{list}%
  \vspace{-0.15\baselineskip}%
}

\begin{table*}[tp]
\centering
\footnotesize
\setlength{\tabcolsep}{5pt}
\renewcommand{\arraystretch}{1.22}
\setlength{\abovecaptionskip}{3pt}
\setlength{\belowcaptionskip}{5pt}
\caption{Recurring ways in which participants used Rasoi during the deployment}
\label{tab:findings-summary}

\begin{tabularx}{\textwidth}{
    >{\raggedright\arraybackslash}p{0.20\textwidth}
    >{\centering\arraybackslash}p{0.09\textwidth}
    >{\raggedright\arraybackslash}X
}
\toprule
\rowcolor{CinderellaBlue}
\textbf{\textcolor{CinderellaBlueText}{Use Case}} &
\textbf{\textcolor{CinderellaBlueText}{Participants}} &
\textbf{\textcolor{CinderellaBlueText}{Description}} \\
\midrule

\cellcolor{CinderellaBluePale}Checking whether changes stayed within scope
& \cellcolor{CinderellaBlueVeryPale}P1, P8
& \cellbullets{
    \item P1 used Rasoi while adding a feature to a command-based developer documentation tool. Rasoi's visual diff showed that the agent's edits stayed only in the expected command-handling path, helping P1 confirm the change was safe to commit.
    \item P8 used Rasoi after asking an agent to add a set-theory visualizer to an educational web app; the diff confirmed that the visualizer and its supporting modules were the only parts that changed to make the new component work.
  } \\
\addlinespace[0.6em]

\cellcolor{CinderellaBluePale}Catching unexpected or missing structural changes
& \cellcolor{CinderellaBlueVeryPale}P4, P7, P9
& \cellbullets{
    \item P4 used Rasoi after a robustness pass on an Android accessibility service; a deletion marker revealed that the agent had removed an audio warning cue, even though the request did not explicitly ask it to remove that feedback.
    \item P7 used Rasoi while extending an ML evaluation pipeline and found that the agent had missed loading/evaluation-script updates, which explained why training appeared to run but evaluation improvements were limited.
    \item P9 asked an agent to modularize a portfolio web app, but Rasoi showed no new or deleted nodes even at a lower abstraction level, making them skeptical even before running the code and seeing that the modularization had broken.
  } \\
\addlinespace[0.6em]

\cellcolor{CinderellaBluePale}Tracing relationships between changes and the larger system
& \cellcolor{CinderellaBlueVeryPale}P2, P3
& \cellbullets{
    \item P2 used Rasoi during a Firebase-to-Supabase migration to see which parts of the system still pointed to Firebase, what still needed to be migrated, and how the remaining service connections affected the larger backend.
    \item P3 used Rasoi on a large, under-documented Fortran simulation codebase to see how data moved between modules, including where variables and matrices were defined, passed, and reused across the system.
  } \\
\addlinespace[0.6em]

\cellcolor{CinderellaBluePale}Rebuilding a usable mental model for review, explanation, or handoff
& \cellcolor{CinderellaBlueVeryPale}P5, P6, P10, P11
& \cellbullets{
    \item P5 used Rasoi to explain the high-level structure of a robotics and 3D simulation codebase to someone unfamiliar with it, using the diagram as a shared reference instead of walking through the code verbally.
    \item P6 used Rasoi while reviewing a collaborator's large pull request, first using the architecture view to understand the overall shape of the change before deciding which parts of the codebase needed closer inspection.
    \item P10 used Rasoi after time away from a web annotation tool to recover how the frontend connected to the backend and regain enough context to continue reviewing changes, while P11 used it to break down a large agentic mortgage platform into a more navigable hierarchy and locate where a small requested change fit within the broader codebase.
  } \\
\bottomrule
\end{tabularx}
\end{table*}

\subsubsection{Visual diffs surfaced unexpected side effects and regressions from agent code generation}
Rasoi also helped participants notice when an agent-made change affected more of the system than they expected. After asking the agent to make the system “more robust,” P4 realized that the agent had also removed an earlier short audio cue used to signal warnings to users (Appendix, Figure~\ref{fig:appendix_visualizations}:C). 
The annotated visual diff showed P4 “not just what got added,” but seeing the “-1” in the diff legend made them think, “what the ..., I didn’t remove anything.” P5 described a similar moment while debugging: “There was one part where I finally understood what the agent actually changed, because I initially didn’t realize that that particular part of the system was even involved in the bigger thing that was going on.” P9 likewise used Rasoi to understand a rendering-related bug in a generated portfolio app, explaining that the visual helped them realize “there even was a rendering engine” shaping the interface and that “oh, my God, it’s the rendering. That’s the issue.” These cases show that visual diffs helped participants notice not only what changed, but also when an agent-made change had introduced unexpected side effects or implicated parts of the system they had not realized were involved.

\begin{figure}[tbp]
    \centering
    \includegraphics[width=\columnwidth]{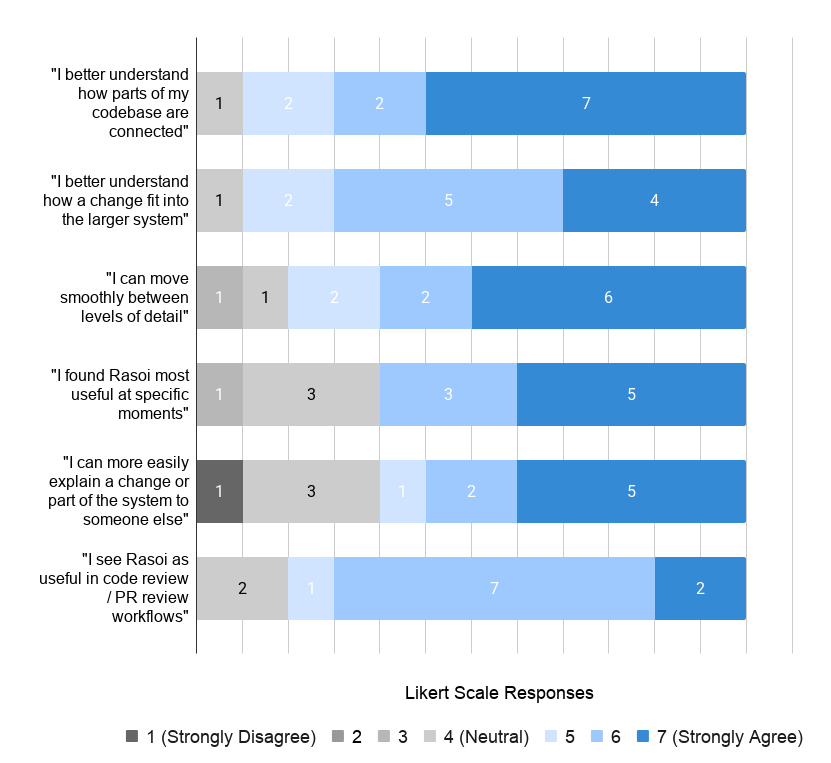}
    \caption{Likert responses to selected post-study statements about Rasoi.}
    \label{fig:survey_likert}
\end{figure}

\subsubsection{Visual diffs helped developers understand how modified parts related to the rest of the codebase}
Participants also used Rasoi to understand how modules that were modified now related to the rest of the codebase. P2 said the visualization helped “organize the architecture better in my mind,” especially when reasoning about whether a component was already connected to several services and a new feature could build on those connections.
The same structural view also helped P2 think about ways to optimize their codebase - they mentioned if “one component is calling an external service twice,” P2 could reason about whether those calls could be reduced.

P3 used Rasoi in a similar way while working in a large codebase inherited from their PI. This was a codebase written primarily in Fortran with limited documentation. What was most useful for them was seeing “how data is transferred between different modules” and, if one module changed, “what impact would that have on everything else?” They also mentioned how without Rasoi, they would have had to keep asking the agent to explain the code, but that “looking at the chat - looking at the graph - is a lot easier.”

P9 also described a similar benefit---they had been rushing code generation and largely ignored the agent’s output during early stages of building, but Rasoi helped them “understand the pieces of the codebase.” This broader structural benefit also appeared in the survey. Eleven of 12 participants selected 5--7 for better understanding how parts of their codebase were connected, and 11 of 12 selected 5--7 for better understanding how a change fit into the larger system (Fig.~\ref{fig:survey_likert}).

\subsection{When do developers need high-level architectural views versus more detailed code-level views to evaluate agent-made changes?}

\subsubsection{High-level views helped participants develop or update a workable mental map of unfamiliar codebases}

Participants often turned to the higher-level views in Rasoi when they did not yet have a usable mental map of the codebase. This was especially true in inherited, under-documented, or quickly evolving systems, where participants knew they needed to make or evaluate changes but did not yet know what parts of the system mattered. P7, who began from a shared codebase they had received from someone else, said the initial visualization “helped me see how the parts were all integrated together” and gave them “a quicker idea” of the system than if they had “just tried to look through the code base myself.” 
Even when they asked Claude for a summary, the agent “did miss some key details,” while the visual gave them enough structural understanding to work from the existing code.

P2 described a similar use while trying to understand a large system. Before using Rasoi, P2 “only knew what was at the context level. [They] knew that there were functions, but .. did not know how they were connected.” With Rasoi, “the animations have been helpful too, because I can actually see what… function is being called, where exactly, by which container,” which revealed “a lot of connections between components” they were “totally not aware of.”

\subsubsection{Developers needed semantically meaningful details in their visuals, not just more details}

When participants asked for more detailed visuals, they wanted details that matched the actual question they were trying to answer rather than denser graphs or more lower-level views, in general. P3 said that Rasoi was still “a bit too macro scale” for some questions. They wanted to be able to trace how changing one component of the code might affect other parts - like how changing the mesh would affect current potential and magnetic field calculations in their system. P2 wanted a different kind of detail. 
The high-level architecture view helped P2 see the overall system as they supervised a data migration from Firebase to Supabase for their system 
(Appendix, Figure ~\ref{fig:appendix_visualizations}: B), 
but they also really needed more precise visibility into which functions or services were still linked to Firebase.
When P1 used the visualization to spot a highlighted module that they did not expect to be modified by the agent, saying “this green thing [the highlighted module]... I don’t know what it is and it shouldn’t have ever been changed.” They added that if they saw “some kind of error now [after running the code],” they could use the visual to point more precisely to the issue: “if I see the green [module], I can just ask the agent, why did you add this thing?” Across these cases, the issue was more about whether the visualization exposed the right abstraction for the participant’s current question. 10 of 12 participants selected 5--7 for the statement that they could move smoothly between high-level views and more detailed views (Fig.~\ref{fig:survey_likert}).

\subsection{Rasoi strengthened developers’ reported confidence in their own understanding of the codebase} 
\begin{figure}[t]
    \centering
    \includegraphics[width=\columnwidth]{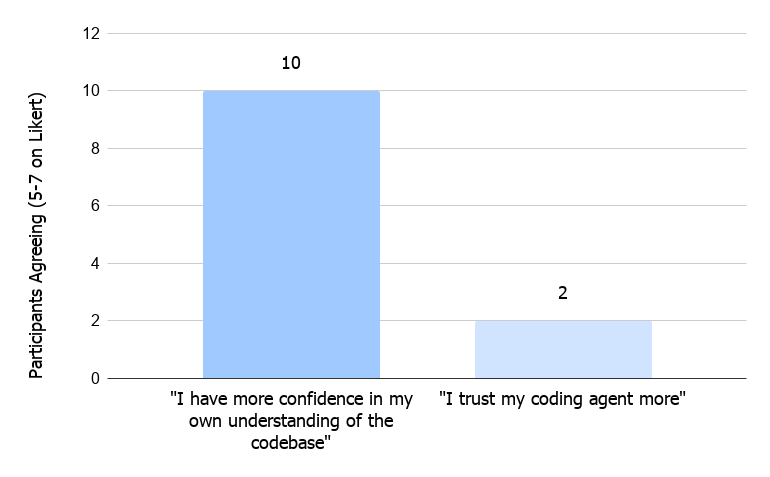}
    \caption{Number of participants selecting 5--7, where 7 indicates ``strongly agree''. More participants reported increased confidence in their own understanding of the codebase than increased trust in the coding agent itself.}
    \label{fig:confidence_trust}
\end{figure}
10 out of 12 participants agreed that Rasoi made them feel more confident in their own understanding of the system, while only 2 agreed that it increased their trust in the coding agent itself (Fig.~\ref{fig:confidence_trust}). P7 said that Rasoi “significantly improved my own understanding of the codebase,” but “did not improve my trust in the coding agent as much,” instead helping them “double-check the coding agent’s changes.” P1 similarly said Rasoi “did improve my confidence in reviewing changes” because it gave them “another way to ensure the changes are aligned with what I wanted, rather than relying only on the agent’s explanation.” P5 also explained that “Rasoi helped my understanding of my code and in turn the system. However, it didn’t make me trust the system more.”

Other participants framed this confidence in terms of control and visibility. P4 said Rasoi helped them understand how changes connected to the larger architecture “so I felt more in control than I would’ve simply using an agent,” while P2 emphasized that Rasoi gave them a fuller picture of a large shared codebase that they “did not have” before.
Rasoi helped developers feel more capable of inspecting, reviewing, and reasoning about agent-made changes, which strengthened their confidence in their own understanding of the codebase. But, it did not necessarily mean they trust their coding agents more.

For a smaller number of participants, these two effects may have been related. P10 noted that “having more confidence in your understanding of the system can also reinforce the idea that you trust your coding agent more to execute a given task.” In other words, once developers used Rasoi to verify that an agent had implemented a change correctly, that successful evaluation could also reinforce their sense that the agent might handle similar tasks well in the future.

\section{Discussion}

\subsection{Transparency in agentic coding should be organized around evaluation moments}
Our findings suggest that developers did not want to monitor Rasoi continuously while coding.
They returned to Rasoi mainly after larger agent-assisted modifications like refactors, failed debugging attempts, or a sequence of changes that made the codebase feel out of sync with what they thought had happened.
In these moments, Rasoi helped them inspect whether a change stayed within expected scope, whether something important had been missed or if an unexpected side effect was introduced.

This points to a different model of transparency for agentic coding, where the goal is not to show everything all the time.
Rasoi was less useful as an ambient tool, and more so used as something that developers turned to when it was time to evaluate what the agent had done.
The key design question is therefore not how to expose everything at all times, but when to surface the structural information needed for evaluation.
In our study, that moment was usually after the agent finished implementing a meaningful change and the developer needed to decide whether the codebase still matched their intent. We also see this in the survey: 8 of 12 participants selected 5--7 for the statement that Rasoi was most useful at specific moments rather than throughout the entire coding process (Fig.~\ref{fig:survey_likert}).

\subsection{Visualizations can externalize tacit knowledge }
\vspace{-0.22em}

In many software teams, there are a small number of people who carry the “big picture” of the codebase in their heads. They know how major modules connect, where important dependencies live, how the system got to its current form, and how changes to one part of the system will affect other parts. That knowledge is often unevenly distributed, hard to document, and difficult for newer collaborators to access. It is also difficult for coding agents to align with when the information remains implicit. Our findings suggest that architectural visualization can make this knowledge more available when it’s needed, rather than forcing teams to only rely on large static documents. Visualizations can make tacit architectural knowledge more understandable and usable during moments when a user is uncertain about the state of the codebase.

Beyond helping people understand a codebase, this kind of externalization can also help experts express the logic behind why a workflow is structured the way it is. Code and comments often capture what a system does at a local level, but they do not always explain higher-level decisions about why certain components exist, why a workflow is organized in a particular sequence, or why one dependency might matter more than another. A visualization creates a place where that reasoning can be attached to the structure itself. Experts could use it to annotate the “why” behind important parts of the system, making procedural knowledge that usually stays in people’s heads more visible to collaborators and, potentially, to coding agents as well. In that sense visualizations may not only help people inspect existing systems, but also provide a way to communicate logic and constraints that future changes should respect. This idea may extend beyond software. Wherever complex procedures depend on tacit expertise, such as medical or bureaucratic workflows, visual representations can help surface the reasoning that usually remains implicit.

\subsection{Different validation questions from users require different semantic views of the same codebase}
Our findings also suggest that developers did not simply want more detail in their visuals; they wanted detail that matched the question they were trying to answer. For example, P7 wanted to trace how a change to a definition in an earlier module would affect downstream equations and computations in their code.
P2 wanted to identify which parts of a migrating system were still connected to Firebase. These are both requests for more detailed views, but not for the same kind of detail. This suggests that future visual support for agent-assisted coding should focus less on making the whole diagram denser and more on foregrounding the structures most relevant to the developer’s current validation task.

\section{Limitations and Future Work}
\textbf{Participant experience.} Participants were experienced developers who regularly used coding agents. We made this choice because Rasoi relies on users being able to reason about code review, software architecture, diff-style representations, and agent-made changes. This gave us access to the review and validation practices the probe was designed to study, but it limits how far our findings extend to novice developers, students, or people newer to agent-assisted workflows.
Future work should study how visual diffs support users with different levels of programming and agent experience, including whether they help experienced developers check and refine existing mental models or help newer users build an initial understanding of unfamiliar codebases. This question also matters for onboarding, since several participants described Rasoi as useful for understanding shared, inherited, or poorly documented codebases before they knew which files, modules, or dependencies mattered (P6, P8, P9, P12).

\textbf{Naturalistic deployment.} Participants used Rasoi on their own codebases rather than completing the same researcher-designed programming task. We chose this because Rasoi is meant to support evaluation during ongoing agent-assisted development. In these moments, developers already have partial mental models of the system, prior expectations about what should change, and uncertainty about whether the agent-made edits still match their intent. This allowed us to observe Rasoi during realistic moments such as refactors, migrations, debugging, code review, and exploratory feature development.
However, this choice also made the deployment harder to standardize and compare across participants. Because each participant worked on a different codebase and task, our study cannot isolate Rasoi's value from alternatives like file-level diffs, agent-generated summaries, or architecture diagrams without versioned diffs. A controlled study with a common task would be valuable for testing these comparisons more directly and for separating the contributions of architecture views, versioned diffs, and drill-down interaction.

\textbf{Probe-stage design.} Rasoi was designed as a probe rather than a finished development tool. The C4 model gave us a stable scaffold for comparing architecture maps across versions, but it was sometimes too heavy for small projects and too generic for specialized tasks, such as tracing Firebase dependencies during a migration or following variables through scientific code. Participants also wanted clearer summaries of what changed, why it changed, and how those changes related to the prompts or tasks they had given the coding agent. Future systems should support more malleable visual diffs that adapt to the dependencies, workflows, or domain-specific concepts developers need to inspect for different tasks. They should also integrate more directly into the places where developers already review and coordinate changes, such as pull requests, code review sessions, CI checks, and IDEs.
\section{Conclusion}

As coding agents increasingly modify software systems, developers need ways to understand how their codebases change across agent-driven iterations. We present Rasoi, a design probe that makes structural change visible through versioned architecture maps, visual diffs, and drill-down across abstraction levels. In a week-long deployment with 12 developers, Rasoi helped participants verify whether agent-made changes stayed within scope, surface unexpected side effects, and understand how modified components related to the rest of the system. Rasoi strengthened participants' confidence in their own understanding of the codebase more consistently than it increased trust in the coding agent itself. These findings suggest that visual support for agent-assisted development should appear at moments of evaluation, match the semantic detail developers need, and treat architectural visibility as a way to support human judgment rather than replace it.

\clearpage
\appendices
\onecolumn
\begingroup
\setlength{\parskip}{0pt}
\setlength{\abovecaptionskip}{2pt}
\setlength{\belowcaptionskip}{0pt}

\section{Rasoi Implementation and Deployment Details}
\label{app:implementation}

Rasoi was implemented as a local web-based design probe for inspecting agent-generated software systems over time. The viewer was built with Vite, React, TypeScript, Material UI, and React Flow, with a lightweight local server for session management, snapshot upload, validation, and optional assistant features. Developers configure Rasoi by copying the provided agent rules into their coding-agent environment. These rules instruct the agent to generate C4-style architecture snapshots using Rasoi's \texttt{DiagramData} prompt, save each version as \texttt{sessions/<session-id>/vN/diagram.json}, and run a validator before viewing. Thus, the target application requires no runtime instrumentation; the agent produces a structured representation directly from the codebase.

Each \texttt{diagram.json} captures the inferred primary user workflow, relevant people or external systems, software containers, internal components, directed data/request-flow edges, stable identifiers, and layout coordinates. These fields allow Rasoi to render elements consistently across versions and highlight additions or modifications as the system evolves.

\vspace{-0.6em}
\section{Additional Participant Diff Visualizations}
\label{app:visualizations}
\vspace{-0.7em}

\begin{figure}[!h]
    \centering

    \setbox0=\hbox{\includegraphics[width=0.68\textwidth]{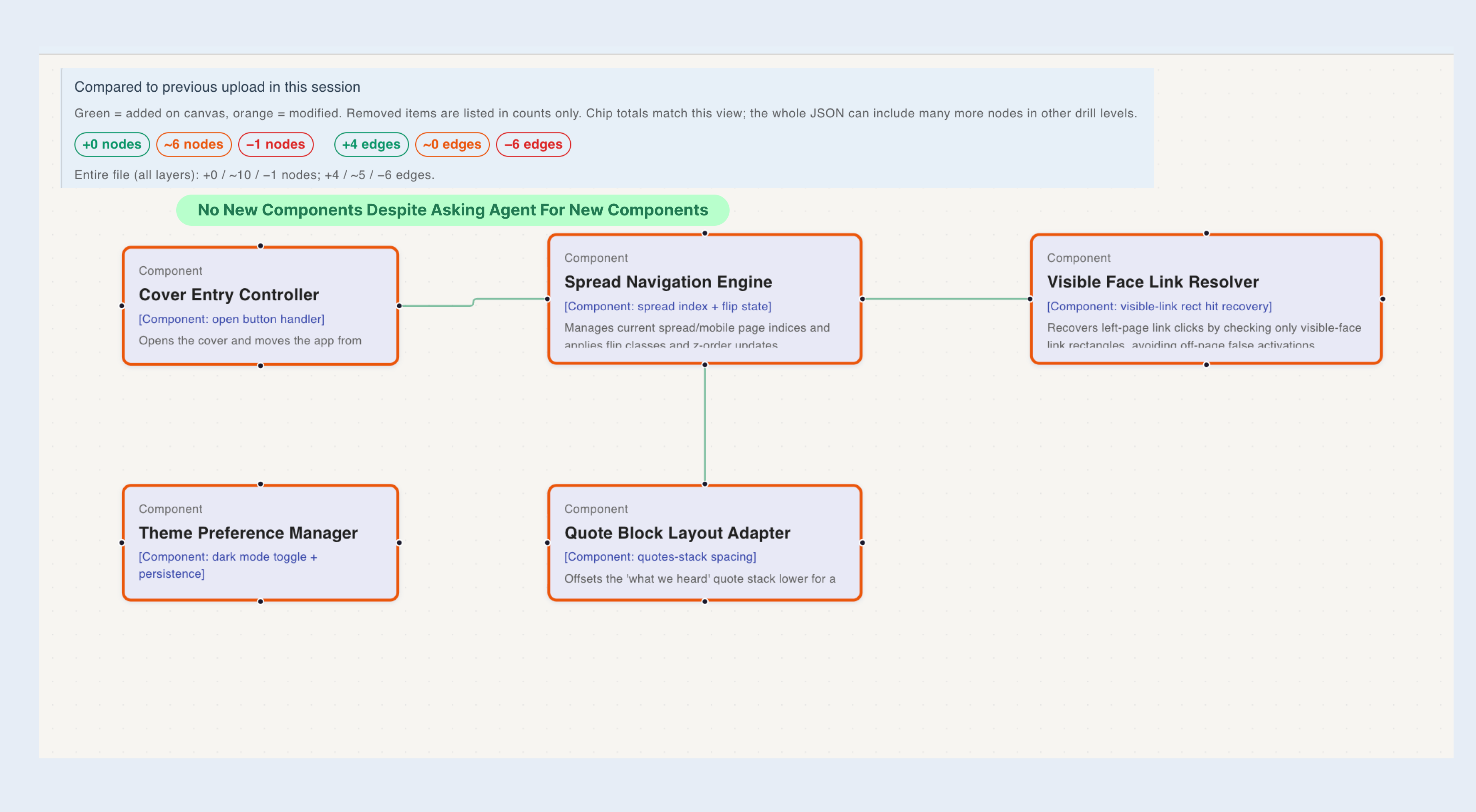}}%
    \makebox[\textwidth][c]{%
        \makebox[0pt][c]{\box0}%
        \makebox[0pt][c]{%
            \raisebox{\dimexpr\ht0-0.4em\relax}{\scriptsize\textbf{A}}%
        }%
        \phantom{\usebox0}%
    }
    \vspace{-0.6em}

    \setbox0=\hbox{\includegraphics[width=0.68\textwidth]{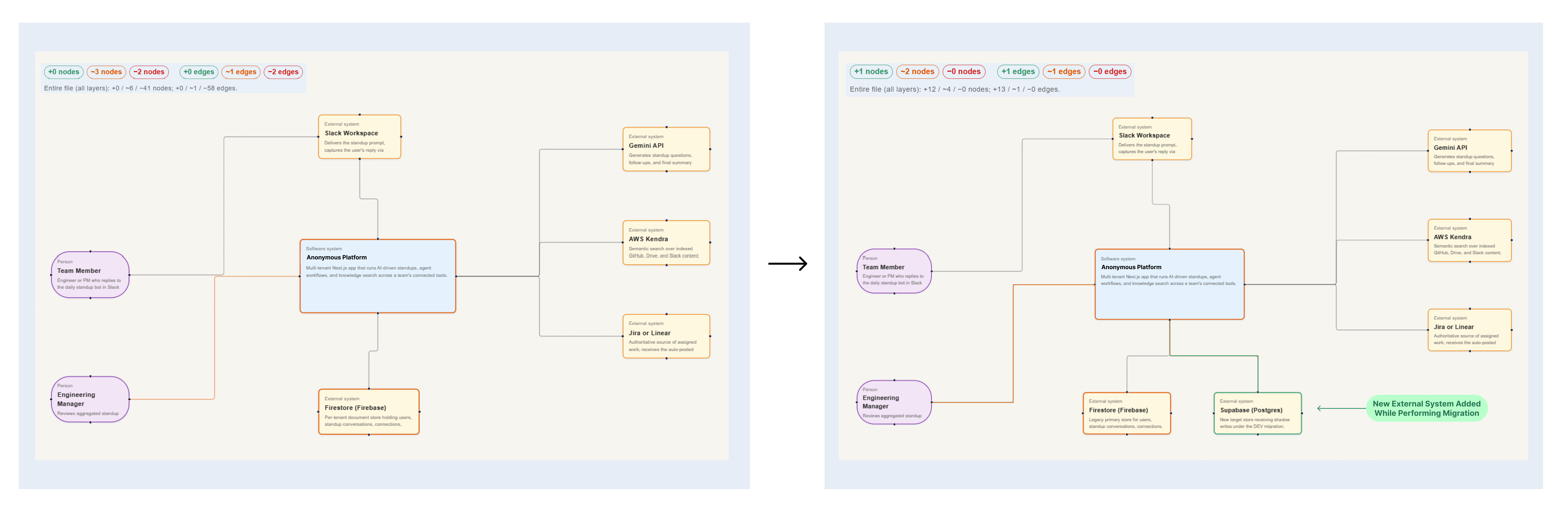}}%
    \makebox[\textwidth][c]{%
        \makebox[0pt][c]{\box0}%
        \makebox[0pt][c]{%
            \raisebox{\dimexpr\ht0-0em\relax}{\scriptsize\textbf{B}}%
        }%
        \phantom{\usebox0}%
    }
    \vspace{-0.6em}

    \setbox0=\hbox{\includegraphics[width=0.68\textwidth]{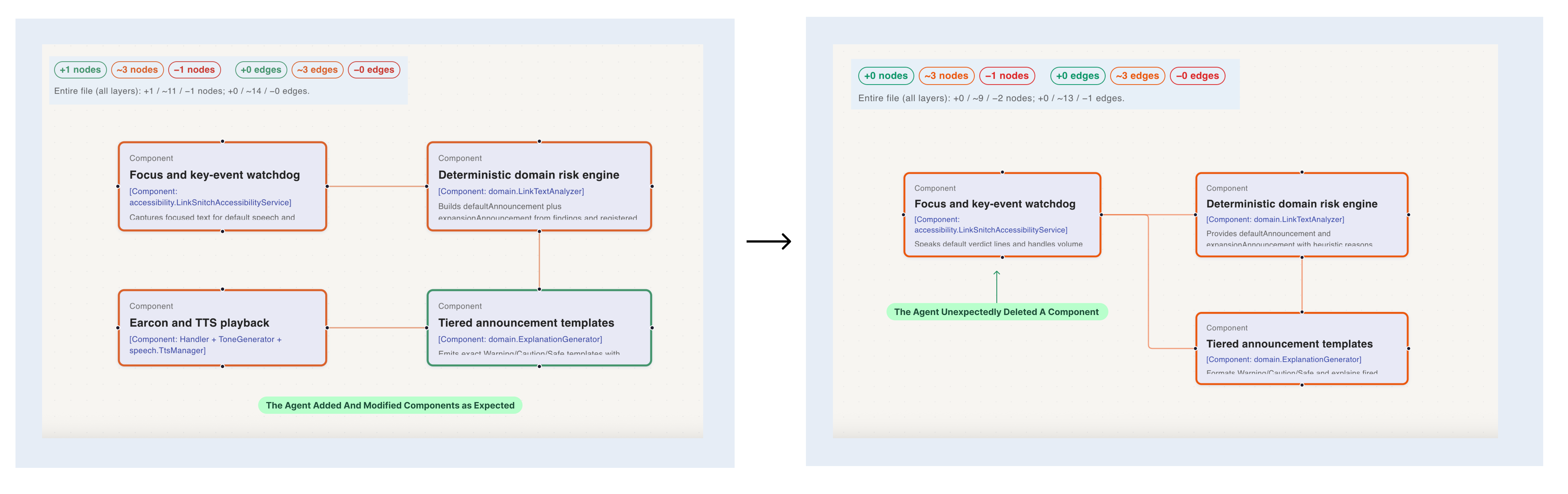}}%
    \makebox[\textwidth][c]{%
        \makebox[0pt][c]{\box0}%
        \makebox[0pt][c]{%
            \raisebox{\dimexpr\ht00\relax}{\scriptsize\textbf{C}}%
        }%
        \phantom{\usebox0}%
    }
    \vspace{-0.4em}

    \caption{Additional Rasoi participant visualizations. A: Rasoi visualization for P9's system. The absence of new or deleted nodes helped reveal that the requested frontend modularization had likely not been applied by the coding agent. B: High-level architecture visualization used by P2 to supervise and understand a Firebase-to-Supabase data migration, including remaining dependencies on both systems. C: Visual diff for P4's system revealing unexpected side effects from agent-generated changes, including removed functionality and previously unnoticed system components implicated in bugs.}
    \label{fig:appendix_visualizations}
    \vspace{-1em}
\end{figure}

\endgroup

\clearpage
\twocolumn
\bibliographystyle{IEEEtran}
\bibliography{sample-base}

\end{document}